\documentclass[10pt, letterpaper, journal]{IEEEtran}
\usepackage{cite}
\usepackage{bm}
\usepackage[cmex10]{amsmath}
\usepackage{graphicx}
\usepackage{url}
\usepackage{color}
\usepackage{amsfonts}
\usepackage{amsmath}
\usepackage{extarrows}
\usepackage{graphicx}
\usepackage{amsfonts}
\usepackage{caption}
\usepackage{algorithm}
\usepackage{algpseudocode}
\usepackage{indentfirst}
\usepackage{caption}
\usepackage{graphics}
\usepackage{epsfig}
\usepackage{soul}
\usepackage{caption}
\usepackage{gensymb}
 \DeclareMathOperator*{\argmax}{argmax}
\DeclareMathOperator*{\argmin}{argmin}

\def\bA{{\mathbf{A}}} \def\bB{{\mathbf{B}}} \def\bC{{\mathbf{C}}} \def\bD{{\mathbf{D}}} 
\def\bF{{\mathbf{F}}} \def\bG{{\mathbf{G}}} \def\bH{{\mathbf{H}}} \def\bI{{\mathbf{I}}} 
    
\def\bP{{\mathbf{P}}} \def\bQ{{\mathbf{Q}}} \def\bR{{\mathbf{R}}}  
\def\bU{{\mathbf{U}}} \def\bV{{\mathbf{V}}} \def\bW{{\mathbf{W}}}  
\def\bZ{{\mathbf{Z}}}
\def\ba{{\mathbf{a}}}    
    
  \def\bm{{\mathbf{m}}} \def\bn{{\mathbf{n}}} 
   \def\bs{{\mathbf{s}}} 
   \def\bx{{\mathbf{x}}} \def\by{{\mathbf{y}}}

\def\C{{\mathbb{C}}}
     
\def\argmin{\mathop{\mathrm{argmin}}}
\def\R{{\mathbb{R}}}
\begin{document}
\title{Millimeter Wave MIMO Channel Tracking Systems }
\author{Jiguang~He$^\dag$,
        Taejoon~Kim$^\dag$, Hadi Ghauch$^\dag$,  Kunpeng Liu$^\star$, Guangjian Wang$^\star$\\
        $^\dag$Department of Electronic Engineering, City University of Hong Kong, Kowloon, Hong Kong\\
        Email: jiguanghe2-c@my.cityu.edu.hk, taejokim@cityu.edu.hk, hghauch@cityu.edu.hk \\
        $^\star$Huawei Technologies, Co. Ltd., Chengdu, China\\  Email:  liukunpeng@huawei.com, wangguangjian@huawei.com\\
        \thanks{The authors at the City University of Hong Kong were supported in part by Huawei
        Technologies.}
        \thanks{H. Ghauch is also affiliated with the School of Electrical Engineering and the ACCESS Linnaeus Center, Royal Institute of
        Technology (KTH), Stockholm, Sweden. Email: ghauch@kth.se }}
 \maketitle

\begin{abstract}
We consider channel/subspace tracking systems for temporally
correlated millimeter wave (e.g., E-band) multiple-input
multiple-output (MIMO) channels. Our focus is given to the tracking
algorithm in the non-line-of-sight (NLoS) environment, where the
transmitter and the receiver are equipped with hybrid analog/digital
precoder and combiner, respectively. In the absence of
straightforward time-correlated channel model in the millimeter wave
MIMO literature, we present a temporal MIMO channel evolution model
for NLoS millimeter wave scenarios. Considering that conventional
MIMO channel tracking algorithms in microwave bands are not directly
applicable, we propose a new channel tracking technique based on
sequentially updating the precoder and combiner. Numerical results
demonstrate the superior channel tracking ability of the proposed
technique over independent sounding approach in the presented
channel model and the spatial channel model (SCM) adopted in 3GPP
specification.
\end{abstract}
\begin{keywords}
Millimeter wave MIMO, temporally correlated channel, channel/subspace
tracking, spatial multiplexing.
\end{keywords}

\section{Introduction}
It is now well projected that conventional cellular systems deployed
in the over-crowded sub-3 GHz frequency bands are not capable to
support the recent exponentially growing data rate demand, even
though advanced throughput boosting multiple-input multiple-output
(MIMO) techniques are employed. Thus, there is gradual movement to
shift the operating frequency of cellular systems from the microwave
spectrum to millimeter wave spectrum (3 GHz to 300 GHz). The large
bandwidth available in the millimeter wave spectrum makes its
application for indoor \cite{Torki2011} and even outdoor
\cite{Sooyoung13} transmission feasible.

However, outdoor millimeter wave channel is challenging. The
millimeter wave propagation suffers from severe path loss and other
environmental obstructions \cite{Marcus05}. The small wavelength of
the millimeter wave (relative to the microwave) ensures that
large-sized arrays can be implemented with a small form factor. As a
result, these outdoor challenges can be overcome by providing
sufficient array gain using large-sized array antennas and analog
beamforming and combining at the base station (BS) and the mobile
station (MS) \cite{Sooyoung13}. Hybrid precoding/combining,
consisting of analog and digital precoders/combiners at the BS/MS,
has been investigated to show close-to-optimal data rate in a cost-
and energy-efficient way
\cite{Alkhateeb13,Sayeed10,Venkateswaran10}. Unlike the conventional
digital precoding/combining technique, the analog precoder/combiner
are composed of groups of phase shifters. This is, the weights of
the analog precocer/combiner have constant-modulus property,
imposing hardware constraints.

Following the parametric channel model from conventional MIMO
systems \cite{Sayeed2002}, the millimeter wave channels can be
modeled using angles of arrival (AoA), angles of departure (AoD) and
propagation path gains
\cite{Torki2011,Sooyoung13,Alkhateeb13,Brady13}. The number of
propagation paths is limited, resulting in a sparse channel, due to
the directional transmission and high attenuation of millimeter
waves. To facilitate the hybrid precoding/combining techniques,
usable channel estimates at the MS and BS are crucial. However,
estimating the channel via conventional training-based techniques is
infeasible and ping-pong based indoor/outdoor channel sounding (or
sampling) techniques have recently been studied in \cite{Torki2011,
Sooyoung13, Alkhateeb13}. The channel sounding probes the channel
using a set of predefined beamforming vectors, i.e., codebook, for a
fixed number of channel uses, and selects the best
beamformer/precoder according to certain criteria.

MIMO systems, in practice, can leverage temporal correlation between
channel realizations to further enhance the system performance. The
previous work in conventional MIMO spatial multiplexing systems
devises various subspace tracking techniques, e.g., stochastic
gradient approach \cite{Banister03JSAC}, geodesic precoding
\cite{Jingnong07}, and differential feedback \cite{Taejoon11}, in
temporally correlated MIMO channels. These techniques adapt the
precoder to the temporal channel correlation statistics and thereby,
enhancing the system performance. Although the extension of these
techniques to the millimeter wave MIMO systems is not
straightforward, it would be possible to devise a useful channel
tracking technique by modifying the previous work (e.g.,
\cite{Banister03JSAC, Jingnong07, Taejoon11}), which is our focus in
this paper.

In this paper, we propose a channel tracking technique for the
millimeter wave MIMO systems employing hybrid precoding/combining
architecture in the temporally correlated MIMO channels. We assume a
time division duplex (TDD) system exploiting channel reciprocity.
The primary challenges we face here are how to tractably model the
channel evolution in the millimeter wave scenario and how to devise
a hybrid precoding/combining algorithm that efficiently tracks the
channel evolution. In the absence of a tractable temporally
correlated millimeter wave channel model, we first present, in this
paper, a temporally correlated MIMO channel evolution model by
modifying the parametric channel models. The proposed channel
tracking algorithm is based on updating each column of the analog
precoder and combiner. We call this approach the mode-by-mode
update, where each mode represents one column of the analog precoder
and combiner. It sequentially adapts the analog precoder and
combiner to the temporal correlation statistics. The digital
precoder and combiner updates follow a conventional MIMO precoder
and combiner design procedure. Simulation results verify that the
proposed channel tracking technique successfully tracks the channel
variations in the presented temporally correlated channel model and
spatial channel model (SCM) \cite{SCM1}.

The rest of the paper is organized as follows. Section II describes
the system model and presents the temporally correlated millimeter
wave MIMO channel model, Section III details the proposed tracking
technique, and the numerical results are presented in Section IV.
Concluding remarks are given in Section V.

\section{System Architecture and Temporally Correlated Channel Model}
In this section, we describe the system model of the millimeter wave
MIMO hybrid precoding/combining systems. We then subsequently
present the temporally correlated millimeter wave MIMO channel
model.
\subsection{System Architecture}
\begin{figure}[t]
  \centering
  \includegraphics[width=9cm]{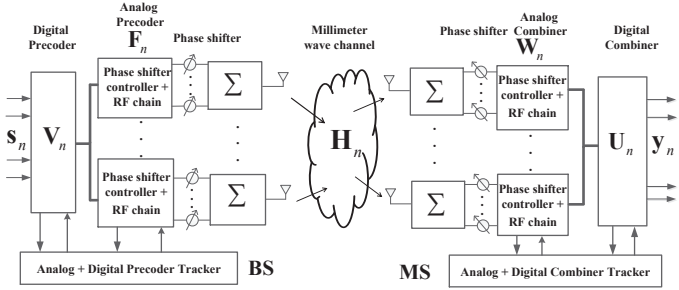}
  \caption{Block diagram of millimeter wave MIMO hybrid precoding/combining and channel tracking system. }\label{system}
\end{figure}
Assuming a narrowband block-fading channel model and $N_s$ data
streams, the output signal\footnote{ A bold capital letter $\bA$ is
a matrix, a bold lower case letter $\ba$ is a vector, and $a$ is a
scalar. $\|\bA\|_F$ is the Frobenius matrix norm of $\bA$.
Superscripts $^T$, $^*$, $^{-1}$, and $^\dagger$ denote the
transpose, Hermitian transpose, inverse, and Moore-Penrose
pseudoinverse operations, respectively. $\bA^{(i,j)}$, $\bA^{(n)}$,
$[\bA]_{:,1:M}$, and $[\bA]_{1:M,:}$ denote the $i$th row and $j$th
column entry of $\bA$, the $n$th column of $\bA$, the first $M$
columns of $\bA$, and the first $M$ rows of $\bA$, respectively.
$\bI_{M}$ is the $M\times M$ identity matrix.
$\bx\sim\mathcal{CN}(\bm,\mathbf{\Upsilon})$ represents the complex
Gaussian random vector with mean $\bm$ and covariance
$\mathbf{\Upsilon}$. $\text{diag}(\bA)$ extracts diagonal elements
of the square matrix $\bA$ and put them into a vector, while
$\text{diag}(\ba)$ stands for a diagonal matrix with $\ba$ on its
diagonal entries. $X\sim\mathcal{U}(a, b)$ denotes the random
variable uniformly distributed on the interval $[a, b]$, and
$E[\cdot]$ denotes the expectation operator.} $\by_n\in
\C^{N_s\times 1}$ in Fig.~\ref{system} at the channel block $n$ in
the downlink is modeled via
\begin{equation}\label{Basic_Rec_Sig}
\by_n = \bU^*_n \bW^*_n \bH_n \bF_n\bV_n\bs_n +\bU_n^* \bW_n^*\bn_n.
\end{equation}
The $\bU_n\in \C^{N_{rf}\times N_s}$ and $\bW_n\in \C^{N_r\times
N_{rf}}$ represent the digital combiner and analog combiner at the
MS, respectively, with $\|\bW_n\bU_n\|_F^2=N_s$. The $N_r$ and
$N_{rf}$, respectively, denote the number of antennas and number of
RF chains at the MS. The $\bH_n\in\C^{N_r\times N_t}$ is the
millimeter wave channel from the BS to MS, where $N_t$ denotes the
number of BS antennas. We assume for simplicity that both BS and MS
have the same number of RF chains. The $\bV_n\in \C^{N_{rf}\times
N_s}$ and $\bF_n\in \C^{N_t\times N_{rf}}$ represent the digital
precoder and analog precoder at the BS, respectively, with
$\|\bF_n\bV_n\|_F^2=N_s$. The $\bs_n\in\C^{N_s \times 1}$ in
\eqref{Basic_Rec_Sig} is the data symbol vector transmitted from the
BS, constrained to have $E[\bs_n\bs_n^*] = \frac{1}{N_s}\bI_{N_s}$,
and $\bn_n\in\C^{N_r\times 1}$ is the additive Gaussian noise
distributed as $\mathcal{CN}(\mathbf{0},\sigma^2\bI_{N_r})$. The
digital precoder $\bV_n$ and combiner $\bU_n$ can adjust both the
phase and amplitude, while only phase control is allowed, as seen
from Fig.~\ref{system}, for the analog precoder $\bF_n$ and combiner
$\bW_n$, i.e.,
\begin{equation}
|\bF_n^{(i,j)}| = \frac{1}{\sqrt{N_t}},\, \text{and} \;
|\bW_n^{(i,j)}| = \frac{1}{\sqrt{N_r}} \ , \forall i,j.\label{c1}
\end{equation}
We assume that the phase of each element in $\bF_n$ and $\bW_n$ is
quantized to $Q$ bits, i.e., $\angle\bF_n^{(i,j)},\,
\angle\bW_n^{(i,j)}~\in\{0,2\pi(\frac{1}{2^Q}),\cdots,
2\pi(\frac{2^Q-1}{2^Q})\}$. In general, we make an assumption that
$N_s \leq N_{rf} \leq \min(N_t, N_r)$.

\subsection{Temporally correlated Millimeter Wave MIMO Channel Model}
Based on the parametric channel model, the millimeter wave MIMO
channel can be reasonably modeled by manipulating the AoAs, AoDs,
and a limited number of propagation path gains, e.g.,
\cite{Torki2011, Sooyoung13, Alkhateeb13, Brady13}. We assume NLoS
scenario \cite{Alkhateeb13}, where the propagation path gains are
modeled as independent and identically distributed (i.i.d.) Gaussian
random variables,
\begin{equation}\label{channel_n}
\bH_n = \sqrt{\frac{N_t
N_r}{L}}\bA_r(\boldsymbol{\theta}_n)\bD_n\bA_t^*(\boldsymbol{\phi}_n),
\end{equation}
where $\bD_n = \text{diag}([\alpha_{n,1},\alpha_{n,L},\cdots,
\alpha_{n,L}]^T)\in\C^{L\times L}$ is the propagation path gain
matrix. The $\alpha_{n,i}$ denotes the gain of $i$th path with
$\alpha_{n,i}\sim \mathcal{CN}(0,1)$ for $i = 1,\cdots,L$, where $L$
is the number of propagation paths. The $\boldsymbol{\phi}_n=
[\phi_{n,1},\phi_{n,2},\cdots, \phi_{n,L}]^T\in\R^{L\times 1}$ and
$\boldsymbol{\theta}_n = [\theta_{n,1},\theta_{n,2},\cdots,
\theta_{n,L}]^T\in\R^{L\times 1}$ in \eqref{channel_n} denote the
AoDs and AoAs of $L$ independent paths. We assume $\theta_{n,i},
\phi_{n,i} \sim \mathcal{U}(-\pi, \pi),\;\forall n,i$. The
$\bA_t(\boldsymbol{\phi}_n)\in\C^{N_t\times L}$ and
$\bA_r(\boldsymbol{\theta}_n)\in\C^{N_r\times L}$ in
\eqref{channel_n}, respectively, represent array response matrices
at the BS and MS,
\begin{equation}
\bA_t(\boldsymbol\phi_n) =
\frac{1}{\sqrt{N_t}}[\ba_t(\phi_{n,1}),\ba_t(\phi_{n,2}),\cdots,\ba_t(\phi_{n,L})]  \label{A_BS}
\end{equation}
and
\begin{equation}
\bA_r(\boldsymbol\theta_n) =
\frac{1}{\sqrt{N_r}}[\ba_r(\theta_{n,1}),\ba_r(\theta_{n,2}),\cdots,\ba_r(\theta_{n,L})]. \label{A_MS}
\end{equation}

With uniform linear array (ULA) assumption, the $\ba_t(\phi_{n,l})$
in \eqref{A_BS} and $\ba_r(\theta_{n,l})$ in \eqref{A_MS} are given
by \small
\begin{equation}
\ba_t(\phi_{n,l}) =
[1,e^{j\frac{2\pi}{\lambda}d\sin(\phi_{n,l})},\cdots,e^{j(N_t-1)\frac{2\pi}{\lambda}d\sin(\phi_{n,l})}]^T
\end{equation}
and
\begin{equation}
\ba_r(\theta_{n,l}) =
[1,e^{j\frac{2\pi}{\lambda}d\sin(\theta_{n,l})},\cdots,e^{j(N_r-1)\frac{2\pi}{\lambda}d\sin(\theta_{n,l})}]^T,
\end{equation}
\normalsize where $\lambda$ is the wavelength and $d$ is the
inter-antenna spacing.

Based on the channel model in \eqref{channel_n}, we now present a
temporal channel evolution model. The evolution of $\bH_n$ from
$n$th channel block to $(n+1)$th channel block follows
\begin{equation}\label{H_n+1}
\bH_{n+1} = \sqrt{\frac{N_t
N_r}{L}}\bA_r(\boldsymbol\theta_{n+1})\bD_{n+1}\bA_t^*(\boldsymbol\phi_{n+1}),
\end{equation}
where
\begin{align}
&\bD_{n+1} = \rho\bD_n+\sqrt{1-\rho^2}\bB_{n+1}, \label{D_N+1}\\
\ &\bA_r(\boldsymbol\theta_{n+1})=\bA_r(\boldsymbol\theta_n+\Delta\boldsymbol\theta_n), \label{A_r_n+1}\\
&\bA_t(\boldsymbol\phi_{n+1}) =
\bA_t(\boldsymbol\phi_n+\Delta\boldsymbol\phi_n). \label{A_t_n+1}
\end{align}
The $\rho=E[\alpha_{n,i} \alpha^*_{n+1,i}]\in [0\; 1]$, $i=1,\ldots,
L$, is the time correlation coefficient, which follows Jakes' model
\cite{Proakis2000} according to $\rho=J_0(2\pi f_D T)$. The
$J_0(\cdot)$ is the zeroth order Bessel function of first kind, and
the $f_D$ and $T$ denote the maximum Doppler frequency and channel
block length, respectively. The $f_D = f_c v/c$, where $f_c$, $v$,
and $c$ represent the carrier frequency (Hz), the speed of the MS
(km/h), and $c=3\times 10^8$ (m/s), respectively. The $\bB_{n+1}$ in
\eqref{D_N+1} is the diagonal matrix with diagonal entries drawn
from $\mathcal{CN}(0,1)$ and independent from $\bD_n$. As shown in
\eqref{D_N+1}, the evolution of the propagation path gains is
modeled as the first order Gauss-Markov process. We assume that
angle variations $\Delta\boldsymbol\theta_n, \Delta\boldsymbol\phi_n
\sim \mathcal{U}(-\delta, \delta)$, where $\delta$ is small.

\section{Subspace Tracking: Algorithm Development}
\begin{figure}[t]
    \centering
  \includegraphics[width=8cm]{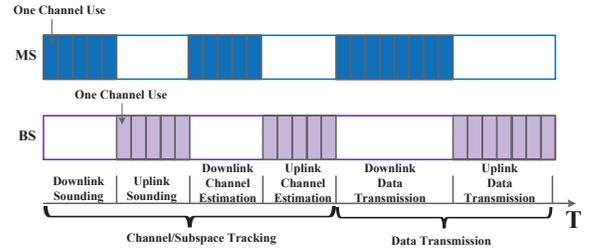}
  \caption{The graphical description of the proposed channel/subspace tracking technique.}\label{scheme}
\end{figure}

In this section, we detail the proposed subspace tracking technique,
consisting of tracking each mode\footnote{Recall that the mode
represents one column of analog precoder and combiner.} of the
analog precoder and combiner one at a time (by generating a
controlled perturbation around the previous mode), then followed by
the digital precoder and combiner update. The motivation of the
mode-by-mode update lies in that we can sound $N_{rf}$ codewords per
channel use by taking the advantage of the hybrid architecture
\cite{Alkhateeb13}. Therefore, the sounding overhead\footnote{The
sounding overhead, here, is the total number of channel uses needed
to design the analog precoder and combiner.} could be significantly
reduced. As shown in Fig. \ref{scheme}, the channel block duration
$T$ is decomposed into $6$ phases - the focus of this work is on the
first $4$ phases.

This section only focuses on the downlink sounding used to update
the analog combiner (Sec \ref{sec3a}), and downlink channel
estimation phase used to update the digital combiner (Sec
\ref{sec3b}). The exactly same reasoning applies for updating
the analog and digital precoders as well, and will thus be omitted,
for simplicity and conciseness.

We summarize the main steps of the algorithm. For each mode, we
construct a codebook by rotating the previous mode using a group of
rotation matrices \cite{Kim2012,Taejoon11}, and then we select the
best codeword that maximizes the received power. After the analog
combiner and precoder design, the digital precoder and combiner
follow using pilot-aided conventional channel estimation of the
effective channel.
\subsection{Downlink Sounding and Analog Combiner Update} \label{sec3a}
The downlink sounding is exploited to update the analog combiner by
utilizing the knowledge of previous analog combiner and time
correlation statistics. Thus, given the analog combiner at instant
$n$, $\bW_{n}$, our aim is to leverage the correlation coefficient
$\rho$ to compute the update for $\bW_{n}$, i.e., $\bW_{n+1}$, done
in a mode-by-mode fashion.

Given that each mode is unit-modulus (i.e., \eqref{c1}) and that
adjacent channel blocks are temporally correlated, it is highly
likely that $\bW_{n}$ and $\bW_{n+1}$ are relatively ``close''.
Thus, we first construct a rotation codebook, $\mathcal{C} =
\{\bC_1,\cdots,\bC_N\}$, $\bC_i \in \C^{N_r \times N_r}$,
$\bC_i^*\bC_i = \bC_i \bC_i^* = \bI_{N_r}$, for $i = 1, \cdots, N$.
By applying each element in $\mathcal{C}$ to $\bW_{n}^{(l)}$, i.e.,
$\bC_k\bW_{n}^{(l)}, \ k = 1, \cdots, N$, we can generate codewords
for the update of $\bW_{n+1}^{(l)}$, which are ``close'' to
$\bW_{n}^{(l)}$. This set of codewords for the update of $l$th mode
can be collected as a matrix,
\begin{equation}\label{gen_codeword}
\widehat{\bW}_{n+1,l} = [\bC_1\bW_{n}^{(l)},\cdots,\bC_N
\bW_{n}^{(l)}], \ l = 1, \cdots, N_{rf},
\end{equation}
where the construction/selection of $\bC_1,\cdots, \bC_N$ is
investigated in Sec \ref{sec3c}. This latter $\widehat{\bW}_{n+1,l}$
is applied to the received signal at the MS antenna to yield
\begin{equation}\label{RecSig}
\bx_{n+1,l} =
\widehat{\bW}_{n+1,l}^*\bH_{n+1}\bF_n\bV_n\overline{\bs} +
\widehat{\bW}_{n+1,l}^*\bn_{n+1},
\end{equation}
where $\bF_n$ and $\bV_n$ are the analog and digital precoder at
channel block $n$, respectively, and $\overline{\bs}$ is the
training vector, e.g., all-ones. Given $\bx_{n+1,l}$, the index of
the optimal codeword is given by
\begin{equation}\label{Selection}
k_{l}^{\star}=\mathop{\argmax\limits}_{1\leq k\leq
N}(|\mathbf{x}_{n+1,l}(k)|^2),
\end{equation}
where $\bx_{n+1,l}(k)$ denotes the $k$th element of $\bx_{n+1,l}$.
Note that the total number of soundings for each mode is
$\lceil\frac{N}{N_{rf}}\rceil$\footnote{ In fact, we can sound
maximum $N_{rf}$ vectors in $ \widehat{\bW}_{n+1,l}$ per every
channel use. The \eqref{RecSig} is formulated by taking all the
sounding as whole for the purpose of simplicity. Since the number of
analog chains is $N_{rf}$, the total number of soundings for each
mode is $\lceil\frac{N}{N_{rf}}\rceil$, which reduces the sounding
overhead $N_{rf}$ times.}. From (14), the $\bW_{n+1}^{(l)}$ is
determined by the $k_{l}^{\star}$th column of
$\widehat{\bW}_{n+1,l}$.

The generation of candidate vectors for the second to last mode
update is slightly different from that of the first mode update. For
the second to last mode, the additional step of subtracting the
effect of the previously updated modes, is needed. This can be done
by employing orthogonal subspace projection,
\begin{equation}\label{substracting}
\widehat{\bW}_{n+1,l} =(\bI_{N_r} - \boldsymbol\Theta
(\boldsymbol\Theta^*  \boldsymbol\Theta  )^{-1}\boldsymbol\Theta^*
)\widehat{\bW}_{n+1,l},\;l\geq2,
\end{equation}
where
\begin{equation}
\boldsymbol\Theta  =
[\bW_{n+1}^{(1)},\cdots,\bW_{n+1}^{(l-1)}]\in\C^{N_r\times (l-1)}.
\end{equation}
As the codebook $\widehat{\bW}_{n+1,l}$ for $l =1,\cdots,N_{rf}$
does not guarantee hardware constraint, one possible approach is to
construct an overcomplete matrix  $\bA_{ \text{can}} \in \C^{N_r
\times N_{\text{can}}}$, $N_{\text{can}} \gg N_r$, with each column
vector in the form of array response vector, i.e., $\bA_{\text{can}}
= [\ba_r(\varphi_1),\cdots,\ba_r(\varphi_{N_{\text{can}}})]$, where
$\varphi_i = -\pi+\frac{(2i+1)\pi}{N_{\text{can}}}\; \text{for} \;i
= 0,\cdots,N_{\text{can}}-1$, and replace each column vector of
$\widehat{\bW}_{n+1,l}$ with one from $\bA_{\text{can}}$ with the
closest distance. Note that the phases of the entries in
$\bA_{\text{can}}$ are also quantized using $Q$ bits. The
replacement of the $m$th column vector of $\widehat{\bW}_{n+1,l}$
can be accomplished by
\begin{equation}\label{closest_distance}
m_l^{\star}=\mathop{\argmin}\limits_{1\leq j\leq
N_{\text{can}}}\sqrt{1-|\widehat{\bW}_{n+1,l}^{(m)^*}\bA_{\text{can}}^{(j)}|^2}
\end{equation}
and
\begin{equation}\label{updating}
\widehat{\bW}_{n+1,l}^{(m)}=\bA_{\text{can}}^{(m_l^{\star})}.
\end{equation}
As a result, the matrix $\widehat{\bW}_{n+1,l}$ in \eqref{updating}
produces nearly orthogonal columns to the previously updated modes,
making $\bW_{n+1}$ approximate unitary matrix.

The selection of the optimal column for $\bW_{n+1}^{(l)}$ for $l\geq 2$ follows the procedures in \eqref{RecSig} and
\eqref{Selection}. The above process is repeated for a number $T_{\text{max}}$ of channel blocks (summarized in Algorithm 1).

\begin{algorithm}[h!]
\caption{Analog Combiner Update} \label{alg1}

\begin{algorithmic}[1]
\State \textbf{Initial Values}: $\bW_0$, $\bF_0$, $\bV_0$, $\bU_0$,
$\rho$, $\bA_{\text{can}}$, $\mathcal{C} = \{\bC_1,\cdots,\bC_N\}$.

\For{$n = 0,\cdots, T_{\text{max}}$}
 \For {$l = 1,\cdots, N_{rf}$}

    \State $\widehat{\bW}_{n+1,l} = [\bC_1\bW_{n}^{(l)},\cdots,\bC_N
\bW_{n}^{(l)}]$; \If {$l \geq 2$} \State $\boldsymbol\Theta  =
[\bW_{n+1}^{(1)},\cdots,\bW_{n+1}^{(l-1)}]$;
    \State $\widehat{\bW}_{n+1,l} =(\bI_{N_r} - \boldsymbol\Theta
(\boldsymbol\Theta^*  \boldsymbol\Theta  )^{-1}\boldsymbol\Theta  ^*
)\widehat{\bW}_{n+1,l}$;

\EndIf
    \State Replace each column of $\widehat{\bW}_{n+1,l}$ according to \eqref{closest_distance}
    \State and \eqref{updating}.
    \State $\bx_{n+1,l} = \widehat{\bW}_{n+1,l}^*\bH_{n+1}\bF_n\bV_n\overline{\bs} +
\widehat{\bW}_{n+1,l}^*\bn_{n+1}$;

    \State $k_l^{\star}=\mathop{\argmax}\limits_{1\leq k\leq
N}(|\mathbf{x}_{n+1,l}(k)|^2)$;
    \State $\bW_{n+1}^{(l)} = \widehat{\bW}_{n+1,l}^{(k_l^{\star})} $;
 \EndFor
  \EndFor

\end{algorithmic}

\end{algorithm}

\subsection{Digital Channel Estimation and Combiner Update} \label{sec3b}
After the analog combiner update, the digital combiner is updated
during  the downlink effective channel estimation (phase 3 in Fig.
\ref{scheme}), based on well-known pilot based channel estimation
schemes. The effective channel
$\bH_{n+1,\text{eff}}\in\C^{N_{rf}\times N_{rf}}$ can be expressed
as,
\begin{equation}\label{EffChan}
\bH_{n+1,\text{eff}}  = \bW_{n+1}^* \bH_{n+1} \bF_{n+1},
\end{equation}
where $\bF_{n+1}$ is the updated analog precoder during the second
phase of the channel tracking technique in Fig. \ref{scheme}. Since
characterizing the statistics of the  effective channel in
\eqref{EffChan} is not tractable, we rather employ a simple least
squares (LS) channel estimation technique \cite{Kay93} and leave any
possible improvements (e.g., using Kalman tracking, particle
filters, etc.) for the future work.

Assume a unitary training signal $\bP\in \C^{N_{rf} \times N_{P}}$,
i.e., $\bP\bP^*=\bI_{N_{rf}}$ with $N_P \geq N_{rf}$ being the
channel training block length\footnote{Similar to the definition of
the sounding overhead, the training overhead is the total number of
channel uses needed to design the digital precoder and combiner.}.
The received signal during the training phase is
\begin{equation}
\bQ_{n+1} =\bH_{n+1,\text{eff}}\bP+\bZ_{n+1},
\end{equation}
where $\bZ_{n+1}\in \C^{N_{rf} \times N_{P}}$ is the additive
Gaussian noise matrix with its elements i.i.d. as
$\mathcal{CN}(0,\sigma^2)$. Then, the LS channel estimate yields
\begin{equation}
\widehat{\bH}_{n+1,\text{eff}} = \bQ_{n+1}\bP^*.
\end{equation}
The singular value decomposition (SVD) is applied to decompose
$\widehat{\bH}_{n+1,\text{eff}}
=\widehat{\bU}_{n+1}\widehat{\mathbf{\Lambda}}_{n+1}\widehat{\bV}_{n+1}^*
$, where $\widehat{\bU}_{n+1}\in\C^{N_{rf}\times N_{rf}}$ and
$\widehat{\bV}_{n+1}\in\C^{N_{rf}\times N_{rf}}$ denote the left and
right singular (and unitary) matrices of
$\widehat{\bH}_{n+1,\text{eff}}$, respectively. The
$\widehat{\mathbf{\Lambda}}_{n+1}\in\mathbb{R}^{N_{rf} \times
N_{rf}}$ is the diagonal matrix $\widehat{\mathbf{\Lambda}}_{n+1} =
\text{diag}([\lambda_1,\cdots,\lambda_{N_{rf}}]^T)$ with
$\lambda_1\geq \lambda_2\geq\cdots\geq \lambda_{N_{rf}}\geq 0$.
Then, $\bU_{n+1} =[\widehat{\bU}_{n+1}]_{:,1:N_s} $ is selected as
the digital combiner at the MS (the same applies for the digital
precoder update in the uplink).
\subsection{Design of Rotation Codebook} \label{sec3c}
This subsection concerns systematic design of the rotation codebook
$\mathcal{C}$. In our approach, the generation of $\mathcal{C}$
utilizes a basis codebook $\mathcal{R}=\{\bR_1, \cdots, \bR_N\}$.
Each codeword $\bR_i\in\C^{N_r\times N_r}$ from $\mathcal{R}$ is
structured such that
\small\begin{equation} \bR_i = \bR^i = \left(
\begin{array}{ccc}
\text{exp}(j\theta_{1})& & \textbf{0}\\
    &\cdots& \\
  \textbf{0} & &\text{exp}(j\theta_{N_{r}})\\
\end{array}
\right)^i ,\ i = 1,\cdots ,N,
\end{equation}
\normalsize $\theta_{k} \in\{0,\cdots,\frac{2(M-1)\pi}{M}\}, \ k =
1,\cdots,N_{r}$, and $ M $ is a positive  prime number such that $M
> N_{r}$. Note that given $\bR$, $\bR^i$ is diagonal, and thus is a unitary matrix (where
$i\in\R$ can be any real number, but in our case, $i = 1, ..., N$).
The basis codebook $\mathcal{R}$ can be reasonably designed by
maximizing the minimum chordal distance between any two codewords in
$\mathcal{R}$, i.e.,
\small\begin{equation} \mathcal{R} =
\underset{\lbrace \widetilde{\mathcal{R}}\rbrace}{\argmax} \left(
\underset{\substack{ 1 \leq i<j \leq N \\ \bR_i, \bR_j \in
\widetilde{\mathcal{R}}}}{\min} \sqrt{N_r-|\textrm{diag}(\bR_i)^*
\textrm{diag}(\bR_j)|^2} \right).
\end{equation}

\normalsize Note that the latter design criterion does not adapt to
the temporal correlation structure of the channel. When time
correlation is presented (i.e., when modes of adjacent channel
blocks are ``close''), the adaptation of the rotation codebook
$\mathcal{R}$ follows
\begin{equation}\label{eq23}
\bG_i = \epsilon\mathbf{I}_{N_r}+\sqrt{1-\epsilon^2}\bR_i,
\;\forall\; i=1,\cdots,N.
\end{equation}
We model the $\epsilon$ as a linear function of $\rho$, i.e.,
$\epsilon=\gamma\rho$ with $0<\gamma \leq 1$. The \eqref{eq23} is
applied to generate points (or perturbations) on a sphere centered
at the point $\epsilon\bI_{N_r}$ with radius $\sqrt{1-\epsilon^2}$.
Since $\bG_i$ is not a unitary matrix, we need to project $\bG_i$
onto the unitary matrix space. The projection can easily be
implemented by normalizing $\bG_i$ such that $\bC_i = \bG_i / \|
\text{diag}(\bG_i) \|_2$, which yields the rotation codebook
$\mathcal{C}$ used to rotate the previous mode in
\eqref{gen_codeword}.
\section{Numerical Results}
The Monte Carlo simulations are carried out to demonstrate the
performance of the proposed tracking scheme. We evaluate the
achievable throughput at channel block $n$, which can be expressed
as
\begin{align}
R_n=&\log_2 \bigg(\text{det} \Big(\mathbf{I}_{N_s}+\frac{1}{\sigma^2 N_s}(\bU_n^*\bW_n^*\bW_n\bU_n)^{-1}\bU_n^*\bW_n^*\bH_n \notag\\
&\bF_n\bV_n\bV_n^*\bF_n^*\bH_n^*\bW_n\bU_n \Big)\bigg).
\end{align}

We set $N_t = N_r = 64$ with $N_{rf} = 4$, i.e., $4$ RF chains, $d =
\frac{\lambda}{2}$, and $L = 4$.

We adopt the SCM \cite{SCM1} to emulate a close-to-real millimeter
wave propagation environment. For the SCM channel emulation, we
choose the `\emph{urban micro}' environment. In order to satisfy the
sparsity property of the millimeter wave MIMO channel, the number of
paths is set to $4$, and the number of subpaths per path is also set
to $4$. We set the powers of the $4$ paths to $[0.8893\; 0.0953\;
0.0107\; 0.0047]$. Note that the default value of subpaths per path
in \cite{SCM1} is $20$, so we just pick the first $4$ subpaths in
our case. The inter-antenna spacing is also set to
$\frac{\lambda}{2}$ at both the BS and MS. We assume the system is
deployed in E-band and set the carrier frequency to $72$ GHz. For
the evaluation, as shown in Fig. 3, several values of velocity for
the MS are considered (e.g., $v=1.0, 3.0, 4.4$ km/h). Notice that we
can not set the $\rho$ parameter in SCM. We keep the default values
of SCM for the remaining parameters.

The maximum number of channel blocks $T_{\text{max}}$ is set to $10$
in our simulation. The number of training sequences for the digital
combiner update is $N_P = 1.5 N_{rf}$, and the SNR (i.e.,
$1/\sigma^2$) is set to $0$ dB. The $\bA_{\text{can}}$ is with the
form of array response matrix (e.g., (4) and (5)) with 200 columns,
i.e., $N_{\text{can}} = 200$. The $Q=4$ bits are used to quantize
the phases of analog combiner and precoder. We assume that the
$\delta$ for angle variations in Section II increases linearly with
$v$, $\delta =3.0^{\circ}, 9.0^{\circ}, 13.2^{\circ}$, corresponding
to the aforementioned $v$ values. For $v=3$~km/h and $f_c = 72$ GHz,
the maximum Doppler frequency is $f_D = 200$ Hz and the coherence
time is roughly $T_c =5$ ms. If we define the channel block length
as $10\%$ (with sufficient coherency) of $T_c$, then the latter is
$0.5$ ms: this value, along with calculated maximum Doppler
frequency, result in the temporal correlation coefficient being
$\rho = 0.9037$, according to Jakes' model \cite{Proakis2000}. We
fix the channel block length to be $0.5$ ms for the computation of
the $\rho$ values and associated velocities. We further assume the
system employs the OFDM technique accommodating a broad bandwidth of
$1$ GHz with oversampling factor $1.1$. Then, provided $1024$ FFT
size, one OFDM symbol duration is $\frac{1024}{1.1\ GHz}\approx 1\
us$, implying the system can resort to around $500$ channel uses
during one channel block. Those values will allow the computation of
the total overhead (sounding overhead plus training overhead)
shortly.

The benchmark scheme (also called independent sounding scheme)
consists of independently updating the analog precoder and combiner
in each channel block without adapting to the channel statistics,
which means the codebook is static across all channel blocks. The
digital precoder and combiner designs follow the same procedures of
our proposed channel tracking technique. As for the initial channel
block of our proposed technique, it follows the independent sounding
scheme. In addition, we fix the total overhead in both schemes to be
similar, i.e., the proposed scheme takes $60$ channel uses while the
benchmark scheme requires $80$ channel uses (for the purpose of
fairness). Thus, the resulting total overhead of the case $\rho =
0.9$ is $12 \%$ for our proposed scheme and $16\%$ for the
independent sounding scheme, making both reasonable.

\begin{figure}[t]
  \centering
  \includegraphics[width=8cm]{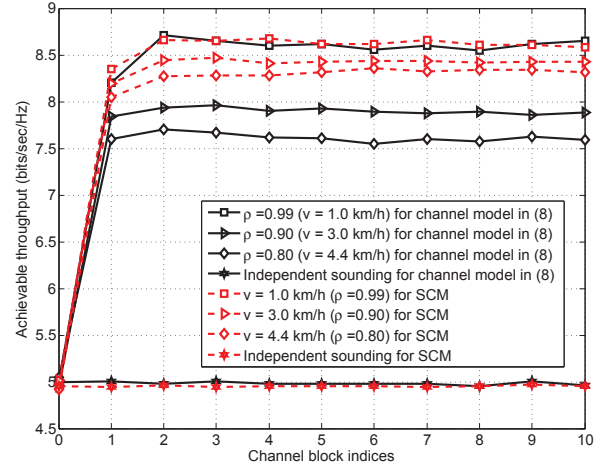}
  \caption{Comparison of achievable throughput of the proposed channel tracking scheme, for both the channel model in (8) and SCM (for several values of $\rho$).}\label{sim1}
\end{figure}

The curves in Fig. \ref{sim1} show that the proposed scheme is able
to track both the temporally correlated millimeter wave MIMO channel
model in \eqref{H_n+1} for different $\rho$ values and SCM for
different velocities. In Fig. 3, we also specify the corresponding
velocity ($\rho$) value for each $\rho$ (velocity) value. The
proposed scheme has similar performance in terms of achievable
throughput in the presented channel model and SCM when $v=1$ km/h.
Despite requiring slightly less total overhead, the proposed
tracking scheme significantly outperforms the independent sounding
scheme.

\section{Conclusion}
We proposed a channel tracking scheme tailored for temporally
correlated NLoS millimeter wave MIMO channels. The proposed scheme
adapts to channel correlation statistics, by constructing
well-designed rotation codebooks, and updates the analog precoder
and combiner in a mode-by-mode fashion. Simulation results
illustrate that the proposed channel tracking scheme outperforms the
independent sounding scheme in the presented parametric channel
evolution model and the SCM.
\bibliographystyle{IEEEtran}
\bibliography{IEEEabrv,Subspace_Tracking_Reference}

\end{document}